# *In-situ* Study of Understanding the Resistive Switching Mechanisms of Nitride-based Memristor Devices

*Di Zhang\*, Rohan Dhall, Matthew M. Schneider, Chengyu Song, Hongyi Dou, Sundar Kunwar, Natanii R. Yazzie, Jim Ciston, Nicholas G. Cucciniello, Pinku Roy, Michael T. Pettes, John Watt, Winson Kuo, Haiyan Wang, Rodney J. McCabe, Aiping Chen\**

Prof. D. Zhang, Dr. S. Kunwar, N. R. Yazzie, Dr. N. G. Cucciniello, Dr. P. Roy, Dr. M. T. Pettes, Dr. J. Watt, Dr. W. Kuo, Dr. A. P. Chen
Center for Integrated Nanotechnologies
Los Alamos National Laboratory
Los Alamos, NM 87545, USA

Prof. D. Zhang
Department of Materials Science and Engineering
University of Texas at Arlington
Arlington, TX 76019, USA

Dr. R. Dhall, C. Song, Dr. J. Ciston
National Center for Electron Microscopy (NCEM), Molecular Foundry
Lawrence Berkeley National Laboratory
Berkeley, CA 94720, USA.

M. M. Schneider, Dr. R. J. McCabe
Materials Science and Technology Division
Los Alamos National Laboratory
Los Alamos, NM 87545, USA

Dr. H. Dou, Prof. H. Wang
School of Materials Engineering
Purdue University
West Lafayette, IN 47907, USA

Corresponding authors: Di Zhang di.zhang@uta.edu; Aiping Chen apchen@lanl.gov

**Abstract**

Interface-type resistive switching (RS) devices with lower operation current and more reliable switching repeatability exhibits great potential in the applications for data storage devices and ultra-low-energy computing. However, the working mechanism of such interface-type RS devices are much less studied compared to that of the filament-type devices, which hinders the design and application of the novel interface-type devices. In this work, we fabricate a metal/TiO$_x$/TiN/Si (001) thin film memristor by using a one-step pulsed laser deposition. *In situ* transmission electron microscopy (TEM) imaging and current-voltage (*I-V*) characteristic demonstrate that the device is switched between high resistive state (HRS) and low resistive state (LRS) in a bipolar fashion with sweeping the applied positive and negative voltages. *In situ* scanning transmission electron microscopy (STEM) experiments with electron energy loss spectroscopy (EELS) reveal that the charged defects (such as oxygen vacancies) can migrate along the intrinsic grain boundaries of TiO$_x$ insulating phase under electric field without forming obvious conductive filaments, resulting in the modulation of Schottky barriers at the metal/semiconductor interfaces. The fundamental insights gained from this study presents a novel perspective on RS processes and opens up new technological opportunities for fabricating ultra-low-energy nitride-based memristive devices.

**1. Introduction**

Resistive switching (RS) in the metal/insulator/metal (MIM) structure has generated broad interests in recent decades among academia and industry for their potential applications in next-generation digital memory and logic devices.[1,2] Resistance random access memory (ReRAM), based on dielectric or semiconducting transition metal oxides (TMO), has shown great potential for high-speed, large-capacity, and non-volatile memory that can be an alternative to flash memory.[3-5] Yet, the critical issue for the ReRAM devices mainly centers on reliability, such as data retention and memory endurance, and the characteristic variations from cell to cell.[2,6] Moreover, the compatibility issue with complementary metal oxide semiconductor (CMOS) processing is another bottleneck for the commercialization of many RS systems.

Besides exploring new material systems and devices towards improved RS characteristics, understanding their RS mechanisms is of equal importance, as it is critical for the optimization and design of next-generation electronic devices. Fundamentally speaking, most RS depends on ionic motions inside the dielectric layer and/or coupled electronic/ionic transport at the metal/semiconductor interfaces of the device.[7] The filament-type RS mechanisms with the formation of conductive filaments (CF) during ion migration and redox processes have been observed in various TMO materials, such as $TiO_2$, $Ta_2O_5$, ZnO, $HfO_2$, and $VO_2$ *etc*. [8-12] The filamentary RS devices usually suffer from undesirable temporal and spatial variations due to the random formation of CFs. In contrast, RS in interface-type RS devices,[13] mainly takes place at the oxide/electrode interfaces.[6,14-16] This type of devices often exhibits lower threshold switching voltage/current and more reliable switching repeatability, thus having great potential in the applications of data storage devices and in-memory computing. Interface-type switching is often coupled with bulk switching. Several different models have been proposed to explain the interface-type switching mechanisms, such as charge (hole or electron) trapping and/or detrapping,[14,17,18] electrochemical migration of oxygen vacancies,[19-21] and Mott transition induced by carriers doped at the interface.[22-24] Some notable examples about the interface-type RS oxides include metal/$Pr_{1-x}Ca_xMnO_3$ (PCMO),[15-17,25] and ferroelectric $BiFeO_3$ and $BaTiO_3$ thin films based memristors.[26-28] However, due to the very limited materials examples and challenges for implementing *in situ* characterization experiments, the RS mechanisms of interface-type devices are much less studied, which undoubtedly hinders its practical applications for memory and logic devices.

*In situ* transmission electron microscopy (TEM) is a promising technique to investigate functional materials and devices down to atomic scale under the influence of a controlled stimulus. The recent development of microelectromechanical systems (MEMS) based TEM technology has significantly expanded the *in situ* TEM capabilities, such as heating, electrical biasing, mechanical testing, and the injection of liquids or gases.[29-35] In this study, we utilize the state-of-the-art *in situ* TEM biasing techniques to study the RS mechanism of different types of memristor devices. Scanning transmission electron microscopy (STEM) with electron energy loss spectroscopy (EELS) are used to characterize and probe the switching mechanism of a nitride film based memristor device integrated on Si (metal/TiO$_x$/TiN/Si). First, *ex situ* electrical characterization is carried out to obtain the current-voltage (*I-V*) characteristics of the bulk device. Next, the TEM lamella of the device is prepared by the focused ion beam (FIB) lift-out technique and then transferred onto the MEMS-based E-chip. *In situ* S/TEM and EELS biasing experiments are conducted to characterize the microstructural evolution and chemical oxidation states variation inside the device during RS process. Furthermore, finite element analysis (FEA) simulations provide theoretical predictions about the electrical field distribution and Joule heating effect within the device during the *in situ* experiment.

2. Results and Discussions
2.1 *I-V* Characteristics of the Bulk Device

As schematically illustrated in Figure 1a, the as-deposited TiN films were annealed in 500 Torr O$_2$ atmosphere at 700 °C for *t* min (*t* = 1, 5, 10) before cooling down to room temperature (RT) in PLD. Figure S1 displays the XRD *θ-2θ* scans of the TiN films after 1-, 5-, and 10-min annealing in 200 mTorr oxygen atmosphere after growth. As expected, the post-annealed films have been partially oxidized and a bilayer of TiO$_x$/TiN was formed. As annealing time increases, the intensity of the epitaxial TiN peak gradually decreases, while the rutile TiO$_2$ peak intensifies, indicating the decrease (increase) of the TiN (TiO$_2$) layer thickness by prolonging the oxidation time.

Figure 1b shows the current-voltage (*I-V*) characteristic loop of a pure TiN film, indicating its high electrical conductivity. In contrast, the Au/TiO$_x$/TiN/P-type Si structures exhibit an obvious RS behavior, as shown in Figure 1c, 1d and S2. The highly repeatable *I-V* loops with ON/OFF ratio close to $10^3$ times indicate the Au/TiO$_x$/TiN structure can be potentially used for resistive memory devices. Compared to the Au/TiN conductive device, it is believed that the oxidized layer TiO$_x$ plays a critical role in its RS behavior. More detailed microstructure characterization and electrical properties analysis are needed to reveal the RS mechanism of the Au/TiO$_x$/TiN/Si memristor device.

## 2.2 *Ex situ* STEM-EELS Analysis

The microstructure of the 5-minute annealed TiN/Si film were first characterized by *ex situ* STEM-EELS. Figure 2a-c show the high-angle annular dark-field (HAADF) STEM and EELS elemental maps for N-*K*, O-*K*, and Ti-*L$_{2,3}$* edges, respectively. The elemental maps (Figure 2b and 2c) show the distinct TiO$_x$/TiN bilayer structure of the device after annealing. To further investigate the elemental distribution and Ti valence states at the TiO$_x$/TiN interface (see inset in Figure 2a), the EELS intensity profiles of the N-*K*, O-*K*, and Ti-*L$_{2,3}$* edges are plotted in Figure 2d-f, respectively. The normalized N-*K* and O-*K* edges EELS intensities displayed in Figure 2g indicates an interfacial layer TiO$_x$N$_y$ formed between the TiO$_x$ and TiN phases. As the $L_{2,3}$ intensity ratio can be correlated with the valence (oxidation) states of 3*d* transition metals due to the electronic occupancy of the $d$ orbitals,[36-38] we calculated the Ti $L_2/L_3$ intensity ratios based on the Ti-$L_{2,3}$ edges (Figure 2f) after removing the background portion of raw energy-loss near-edge structure (ELNES) spectra using power-law and Hartree-Slater cross-section step functions.[37-40] Figure 2h shows the Ti $L_2/L_3$ intensity ratios extracted from the Ti-$L_{2,3}$ EEL spectra (from 1 to 8) at the TiO$_x$/TiN interface. It is clear that the $L_2/L_3$ ratio increases from TiN (Ti$^{3+}$) to TiO$_2$ (Ti$^{4+}$), while the transitional region with a mixture of Ti$^{3+}$ and Ti$^{4+}$ valence states indicate the intermediate TiO$_x$N$_y$ phase. Based on the normalized N-$K$ and O-$K$ edges EELS intensities (Figure 2g) and calculated $L_2/L_3$ intensities ratios (Figure 2h), the intermediate TiO$_x$N$_y$ layer thickness is estimated about 6-8 nm, as the schematics shown in Figure 2i. For the TiN/Si (bottom) and Pt/TiO$_2$ (top) interfaces, the EELS intensity profiles of the N-$K$, O-$K$, and Ti-$L_{2,3}$ edges are plotted in Figure S3. There is no peak shift or Ti-$L_{2,3}$ ratio change observed, indicating the sharp TiN/Si (bottom) and Pt/TiO$_2$ (top) interfaces. Figure S4 shows the HAADF-STEM images and EDS maps of the same series of TiN/Si (001) films after 1- and 10-min annealing in O$_2$ atmosphere.

### 2.3 *In situ I-V* Characteristic and TEM imaging

Next, *in situ* S/TEM biasing experiments were carried out to investigate the RS mechanism. Figure 3a depicts the configuration of the MEMS-based E-chip and the wiring connection for the *in situ* biasing experiment. Figure 3b shows the FIB image of the TEM lamella mounted and welded on the four-point electrical chip. The two narrow trenches on the lamella were cut by Ga$^+$ ion beam to create a vertical memristor device. Figure 3c displays the defined positive and/or negative bias applied on the nanodevice. During the biasing experiment, a compliance current of 40 μA was set to protect the device from thermal breakdown. Figure 3d shows the characteristic *I-V* curve for the Pt(top electrode)/TiO$_x$/TiN(bottom electrode) device by sweeping the applied voltage 0 → +0.4 V → 0 → -0.4 V. Figure 3e plots the *I-V* hysteresis loop in the log-scale. It is clear that the device was switched between the HRS and LRS in a bipolar "counterclockwise-counterclockwise" (CC-CC) manner by sweeping the positive and negative voltages.[41] This switchable diode effect (SDE) *I-V* characteristic has been observed in some metal/ferroelectric/metal (M/FE/M) memristors, such as Ca-doped BFO,[26] Sr-doped BTO,[27,28] and doped PZT.[42] The high concentrations of dopants/acceptors in the ferroelectric layers are believed to play a critical role in the SDE for those devices.[43,44] It should be also noted that such a CC-CC rotation sequence excludes the filament-type switching mechanism as it often shows CC-C or C-CC type rotation sequences by forming and rupturing the conducting filaments at "ON" and "OFF" states.[41,45]

Figure 4a shows the TEM images of the Pt/TiO$_x$/TiN/Si memristor which were acquired at different magnitude of electrical biases applied on the device during the *in situ* biasing experiment. From images A to F, the applied DC bias is labeled on the top right and the measured electrical current and resistance values are labeled on the top left in each image. Figure 4b plots the profiles of applied bias (V), current (µA), and resistance (MΩ) values during the HRS → LRS → HRS switching process, where the six data points (A to F) corresponding to the individual images in Figure 4a are labeled on the three profiles. TEM image series taken at other locations of the specimen including the HRS → LRS or LRS → HRS transitions of the device during switching are recorded as several short movies shown in the Supplementary Materials. Based on the TEM images shown in Figure 4a and Movies S1-S4, no filamentary-like structure was observed in either HRS or LRS for the device, indicating that the RS process may not rely on the formation of conductive filaments. More importantly, it is found that the *I-V* characteristic obtained in the *in situ* TEM biasing experiment (Figure 3d) is considerably different from that of the bulk device (Figure 1c,d). This dramatic change can be explained by the geometry difference between the bulk device and FIB lamella. As the thickness of the TEM lamella is less than 100 nm, the local defects and charge carriers' accumulation or mitigation at the metal/semiconductor interfaces play a more dominant role in controlling the overall resistance for the device. More detailed discussions are shown in the later section of the paper.

### 2.4 *In situ* STEM-EELS Analysis

To further understand the RS mechanism of the device, *in situ* STEM-EELS experiment was carried out during the device switching. Figure 5a shows the HAADF-STEM image of the Pt/TiO$_x$/TiN/Si structure. It is clear that the oxidation layer of TiO$_x$ shows a polycrystalline structure with a large number of structural defects or nanohole-like structures, which might be reservoirs for various charge species such as oxygen vacancies ($V_{\ddot{o}}$), oxygen ions (O$^{2-}$), and/or titanium cations (Ti$^{3+}$, Ti$^{4+}$) that can be migrating under electric fields. To trace the distribution and migration of atomic defects such as oxygen vacancies and other charge species inside the TiO$_x$ layer during RS, we conducted *in situ* EELS spectrum imaging (SI) using Gatan GIF K3 detector on the device at its pristine state (HRS) and after applying +/- biases (LRS). From the EELS SI datasets, we extracted the EELS spectra at different locations and then calculated the averaged Ti $L_2$/$L_3$ intensity ratios after the spectra background removal using power-law and Hartree-Slater cross-section step functions.[37-40] A manually written Python script was used to fit the extracted spectra to obtain Ti $L_2$/$L_3$ intensity ratios. Figure 5b shows the color maps of calculated Ti $L_2$/$L_3$ intensity ratios in Ti $L_{2,3}$ EELS for the TiO$_x$ phase region (red rectangle in Figure 5a) at unbiased (HRS), negative and positive biased (LRS) conditions. For the unbiased map (Figure 5b, left), the $L_2$/$L_3$ ratios in the majority part of TiO$_x$ are close (from 1.34 to 1.38) except for the Pt/TiO$_x$ top interface region with the $L_2$/$L_3$ ratio less than 1.32 which could be due to the oxygen deficiency caused by ion beam during the FIB lamella milling process. When the device was negatively biased and switched to LRS, the $L_2/L_3$ ratio map (Figure 5b, middle) shows an accumulation of $Ti^{4+}$ ions near the TiN bottom electrode. Simultaneously, the interface region near the top Pt electrode, shows a reduction of $L_2/L_3$ ratios which implies a local accumulation of oxygen vacancies that reduce the average valence state of Ti. By flipping the device to a positive bias configuration, the EELS map (Figure 5b, right) shows an opposite trend of the $L_2/L_3$ ratios, where oxygen vacancies migrate towards the bottom TiN electrode. Figure 5c plots the line profiles of the averaged Ti $L_2/L_3$ intensity ratios for each row of the EELS map shown in Figure 5b. It can be seen that the location of the maximum $L_2/L_3$ ratios are switched as we flipped the electric bias direction applied on the memristor device. Our *in situ* EELS Ti $L_2/L_3$ ratio maps provide the direct evidence for the electric field induced migration of oxygen vacancies within the $TiO_x$ films during the RS process. We find that the formation of a conducting filament is not necessary to enable switching the device from HRS to LRS, as this change can be induced by the redistribution of oxygen vacancies at the metal/semiconductor contact interfaces.

**2.5 RS Mechanism Discussion**

The *in situ I-V* characteristic loop shown in Figure 3d and 3e indicates that the lamella device is working as a bipolar rectifier in which the device can be switched from HRS to LRS at a low threshold voltage ~ ± 0.4 V. In general, the rectifying behavior in semiconductors comes from a p-n junction or a metal-semiconductor Schottky junction due to the existence of an energy barrier at the interface.[6,13,41] We therefore hypothesize that the charge species (*i.e.* oxygen vacancies, oxygen ions, and free electrons *etc.*) created to maintain the charge neutrality are critical to understand the conduction behavior for the device. Figure 6a illustrates the charge neutrality state of the device at unbiased condition. Upon applying a vertical electric field ($E$) to the device, the mobile positive charges (*e.g.* oxygen vacancies) and negative charges (*e.g.* oxygen ions, electrons) can move through the device to find a new thermodynamic equilibrium. For instance, under negative bias ($E$-) condition, the oxygen vacancies or any positive charges migrate and pile up near the top interface, whereas the negatively charged species move down and accumulate close to the bottom interface. Consequently, the local imbalance in positive (donors) and negative (acceptors) ions create an electric field driven n-p junction (shown in Figure 6b). Similarly, under positive bias ($E$+) the insulating layer forms a p-n junction, as illustrated in Figure 6c.

The Schottky-like barriers at the metal/film interfaces play a critical role in explaining the diode-like *I-V* characteristic. [13,41] The Schottky barrier height and width can be determined by the difference between the metal work function ($\phi$) and the semiconductor electron affinity ($\chi$). Figure S5 shows the energy band diagrams of Pt, TiO$_2$, and TiN before being contacted. It is noted that the work function of Pt is ~ 5.3-5.65 eV [46-48] and the electron affinity of TiN is ~ 4.7-5.3 eV,[49-51] while the electron affinity of TiO$_2$ is ~ 4.2-4.4 eV.[52-54] When TiO$_2$ and Pt or TiN are physically contacted, some of the electrons in TiO$_2$ move spontaneously into Pt or TiN due to the higher Fermi level of TiO$_2$ than that of Pt or TiN, resulting in the formation of Schottky barrier ($\phi_B$) and depletion region ($W_d$) at both Pt/TiO$_2$ and TiO$_2$/TiN interfaces, as illustrated in Figure 6d. For a pristine state device under zero or ultra-low electric field, due to the two opposite Schottky barriers at the top and bottom interfaces, the oxygen vacancies as donor impurities cannot move easily, thus the Pt/TiO$_2$/TiN device exhibits a very high resistive state no matter the applied voltage is positive or negative. When a sufficient high negative bias ($E$-) is applied onto the device, the positively charged oxygen vacancies move to the Pt/TiO$_2$ upper interface (as the blue shadow area shown in Figure 6e). The depolarization field drives the released electrons to neutralize the positive bound charges at the interface and a narrow electron region would be formed near the interface with positive bound charges, resulting in the downward-bending of the band at the upper (right) interface.[55-57] As a result, the Pt/TiO$_2$ upper interface becomes Ohmic contact under the large *E*-. On the other hand, the negative charges migrate to the TiN/TiO$_2$ bottom interface simultaneously, leading to the upward-bending of the band structures (Figure 6e). Consequently, at the lower interface between TiO$_2$ and TiN, the built-in potential ($\phi'_{bi}$) increases and the depletion region ($W'_d$) becomes wider, accompanied with an enhanced Schottky barrier ($\phi'_{B(bot)}$). Therefore, the modulation of upper and lower Schottky interfaces leads the device to behave like a forward diode under negative biased condition (Figure 6e), where the enhanced Schottky-like barrier at TiO$_2$/TiN interface plays a dominant role in the conductive characteristics. Similarly, when a sufficient positive bias (*E*+) applied, the device shows opposite modulations to the band structures and Schottky-like interfaces, resulting in a reverse diode *I-V* behavior as shown in Figure 6f.

Another important question is that why the measured current was abruptly enhanced at a certain threshold voltage despite the fact that, according to Ohmic conduction theory, the charge carriers concentration and device resistance should be continuously varied by sweeping the applied voltage. C.H. Yang *et al.* has explained this phenomenon by using an electronic localization-delocalization transition mechanism via band-filling control.[26] When the Fermi energy ($\varepsilon_F$), which is controlled by the oxygen vacancy concentration ($n_{V_o}$), passes across the mobility edge ($\varepsilon_c$), an electronically conducting state appears locally. Moreover, the Schottky junction current can be described by thermionic emission theory:[14]

$$J_{th} = SA^*T^2\exp(-\frac{q\phi_B}{k_BT})[\exp(\frac{qV}{k_BT})-1]$$

Where *S* is the junction area, *A\** is the Richardson constant, *T* is the temperature, $\phi_B$ is the Schottky barrier height. As the device is switched from HRS to LRS, the tunneling current increases almost 10$^3$ times, resulting in an enhancement of the Joule heating effect. In order to clarify the electro-thermal correlation inside the device during RS process, we conducted finite element simulations using COMSOL Multiphysics (See Experimental Section). The 3D geometry model shown in Figure S6a was retrieved based on the dimension of the TEM lamella. Figure S6b and S6c show the simulated electric potential distribution map with streamline curves indicating the local electric field directions inside the device at HRS and LRS, respectively. The magnified images shown in Figure S6d and 6e show that the electric field distribution is vertical at both Pt/$TiO_2$ and $TiO_2$/TiN interfaces, while the TiN conductive layer as bottom electrode conducting the currents horizontally throughout the device. At HRS, the $TiO_2$ layer behaves like an insulator and blocks the majority of current. When the device is switched to LRS, the electric current increases $10^3$ times and the electric potential varies significantly, as shown in Figure S6e. The electro-thermal simulation results are shown in Figure S7. At HRS, the electric current is at $10^{-8}$ A scale, so the temperature distribution inside the device is quite uniform and close to RT (Figure S7a). At LRS, however, the device current increases to $10^{-5}$ A scale, leading to the enhanced Joule heating effect and substantial increase of the device temperature to more than 500 °C or even higher (see Table S1), which in turn can facilitate the thermionic emission and tunneling current at the Schottky interfaces and set the device to a stable "ON" (LRS) state. When the applied bias drops down to close to zero, the charge carrier's distribution in the $TiO_2$ switching layer becomes homogeneous again and two Schottky barriers are reconfigured at the Pt/$TiO_2$ and $TiO_2$/TiN interfaces, leading the device to be reset to "OFF" (HRS).

3. **Conclusion**

We fabricate metal/$TiO_x$/TiN/Si (001) memristor devices using one-step PLD. The current-voltage (*I-V*) characteristic and RS mechanisms of the device are systematically explored by *in situ* TEM and STEM-EELS biasing experiments. *In situ* EELS Ti $L_2/L_3$ ratio maps clarify the Ti valence states variation inside the device as a result of oxygen vacancies migration under negative and positive electric fields. The RS mechanism discussed is based on the interplay of ionic charges and electronic conduction, namely the electric field-driven n-p and p-n junction formation and Schottky-like barriers modulations at the upper and lower metal/semiconductor interfaces, which result in the bipolar diode-like *I-V* characteristic of the device. The study presents a new perspective of the interface-dominated RS process and opens up technological opportunities for fabricating novel nitride-based memristive devices that are compatible with CMOS processes.

## 4. Experimental Section
### 4.1 Growth of Thin Films

The TiN thin films were deposited on a P-type doped Si (001) substrate using a TiN target by a pulsed laser deposition (PLD, KrF excimer laser, $\lambda = 248$ nm) system. The laser fluence for all depositions was maintained at ~2.0 J/cm$^2$. The deposition process was carried out under high vacuum conditions (~$1\times10^{-6}$ Torr) at a temperature of 700 °C. A total of 15,000 laser pulses with a frequency of 10 Hz were utilized during the TiN film growth. After deposition, pure $O_2$ was inflowed into the PLD chamber to maintain a partial pressure at 500 Torr for 1-, 5-, and 10-min sample annealing, respectively. After that, all the films were cooled down to RT under 500 Torr $O_2$ atmosphere. To confirm the phase and crystallinity of the as-deposited pure TiN and oxidized $TiO_xN_y$/TiN films, X-ray diffraction (XRD) measurement was performed using a high-resolution four-circle X-ray diffractometer (Smartlab, Rigaku Inc.) with monochromatic Cu K$_{\alpha 1}$ ($\lambda = 1.5406$ Å) radiation source.

### 4.2 Electrical Characterization

For the top electrodes, circular Au disks of 100 um diameter were deposited using e-beam evaporation through predesigned metal mask at 140 °C on $TiO_xN_y$/TiN/Si samples. Current-voltage (*I-V*) characteristics were measured with Keithley 2636B SMUs under DC voltage sweep mode. All electrical measurements were performed at ambient condition (room temperature and relative humidity of ~ 12 %) by applying bias on Au disks (top electrode) and grounding the TiN (bottom electrode).

### 4.3 Scanning Transmission Electron Microscopy and Spectroscopy

Cross-sectional TEM samples were prepared by a focused ion beam (FIB) lift-out process on a FEI Helios 600 dual beam SEM/FIB with $Ga^+$ ion source operated at 30 kV. The TEM lamellae were first transferred onto MEMS-based FIB-Optimized E-chips (Protochips Inc. Fusion Select). The TEM lamellae thinning was performed on a Scios 2 Dual Beam SEM/FIB with a Leica VCT cryogenic stage installed. To preserve the structural integrity of the beam sensitive materials and reduce artifacts inclusion the sample was cooled to -150 °C and $Ga^+$ FIB milling was performed with a reduced accelerating voltage of 16 keV. Final thinning and cleaning were conducted using a 5 kV beam at 16 pA followed by 2 kV at 9 pA to remove the ion-damaged amorphous layer on the sample surface. A Protochips Fusion Select double-tilt holder and Keithley 2635B SourceMeter were used to run the *in situ* S/TEM biasing experiment. The drift-correction software AXON (Protochips Inc.) is installed on the TEM computer to acquire drift-corrected TEM images series and videos during the *in situ* biasing experiment. The *in situ* TEM images and energy dispersive X-ray spectroscopy (EDS) mappings were acquired on an image-corrected FEI Titan 80-300 S/TEM operated at 300 keV at LANL. HAADF-STEM and EELS data were collected on a double-corrected FEI Titan 80-300 transmission electron microscope (TEAM 1) equipped with a Gatan Continuum energy filter at National Center for Electron Microscopy (NCEM, Molecular Foundry, LBNL). HAADF-STEM images were collected at 300kV with a convergence semi-angle of 30 mrad and a HAADF collection angle of 55 to 200 mrad. Electron energy loss spectra were recorded on a post-GIF (Gatan energy filter) K3 detector in electron counting mode with a dispersion of 0.18 eV per pixel. The spectrometer was operated in dual-EELS mode to record both the elastic (zero-loss) and core-loss portion of the spectrum. To reduce random noise in the EELS spectra, a principal component analysis (PCA) was applied to every EELS spectrum followed by the background removal by a power-law function.

### 4.4 Finite Elements Simulation

COMSOL Multiphysics *AC/DC Module* with *Electric Currents (ec)* and *Joule Heating physics interfaces* were used for the electric field and Joule heating effect simulation. The simulated geometry was retrieved based on the 3D dimension of the FIB lamellae during the *in situ* STEM-EELS biasing experiment.

## Supporting Information

Supporting Information is available from the Wiley Online Library or from the author.

## Acknowledgements

The work at Los Alamos National Laboratory was supported by the NNSA's Laboratory Directed Research and Development Program, and was performed, in part, at the Center for Integrated Nanotechnologies, an Office of Science User Facility operated for the U.S. Department of Energy (DOE) Office of Science by Los Alamos National Laboratory (contract 89233218CNA000001). Los Alamos National Laboratory, an affirmative action equal opportunity employer, is managed by Triad National Security, LLC for the U.S. Department of Energy's NNSA, under Contract No. 89233218CNA000001. Work at the Molecular Foundry was supported by the Office of Science, Office of Basic Energy Sciences, of the U.S. Department of Energy under Contract No. DE-AC02-05CH11231. The work at Purdue University was funded by the U.S. National Science Foundation (DMR-2016453).

**Competing interests**

The authors declare no conflict of interest.

**Data and materials availability**

**References**

[1]     K. Sun, J. Chen, X. Yan, *Adv Funct Mater* **2021**, *31*, 2006773.

[2]     M. Lanza, H. P. Wong, E. Pop, D. Ielmini, D. Strukov, B. C. Regan, L. Larcher, M. A. Villena, J. J. Yang, L. Goux, *Adv Electron Mater* **2019**, *5*, 1800143.

[3]     G. I. Meijer, *Science (1979)* **2008**, *319*, 1625.

[4]     R. Waser, R. Dittmann, G. Staikov, K. Szot, *Adv Mater* **2009**, *21*, 2632.

[5]     H. Y. Lee, P. S. Chen, T. Y. Wu, Y. S. Chen, C. C. Wang, P. J. Tzeng, C. H. Lin, F. Chen, C. H. Lien, M.-J. Tsai, in *2008 IEEE International Electron Devices Meeting*, IEEE, **2008**, pp. 1–4.

[6]     A. Sawa, *Materials today* **2008**, *11*, 28.

[7]     Y. Yang, W. Lu, *Nanoscale* **2013**, *5*, 10076.

[8]     D.-H. Kwon, K. M. Kim, J. H. Jang, J. M. Jeon, M. H. Lee, G. H. Kim, X.-S. Li, G.-S. Park, B. Lee, S. Han, *Nat Nanotechnol* **2010**, *5*, 148.

[9]     G.-S. Park, Y. B. Kim, S. Y. Park, X. S. Li, S. Heo, M.-J. Lee, M. Chang, J. H. Kwon, M. Kim, U.-I. Chung, *Nat Commun* **2013**, *4*, 2382.

[10]    J.-Y. Chen, C.-L. Hsin, C.-W. Huang, C.-H. Chiu, Y.-T. Huang, S.-J. Lin, W.-W. Wu, L.-J. Chen, *Nano Lett* **2013**, *13*, 3671.

[11]    Y. Zhang, G.-Q. Mao, X. Zhao, Y. Li, M. Zhang, Z. Wu, W. Wu, H. Sun, Y. Guo, L. Wang, *Nat Commun* **2021**, *12*, 7232.

[12]   S. Cheng, M.-H. Lee, X. Li, L. Fratino, F. Tesler, M.-G. Han, J. Del Valle, R. C. Dynes, M. J. Rozenberg, I. K. Schuller, *Proceedings of the National Academy of Sciences* **2021**, *118*, e2013676118.

[13]   S. Bagdzevicius, K. Maas, M. Boudard, M. Burriel, *Resistive Switching: Oxide Materials, Mechanisms, Devices and Operations* **2022**, 235.

[14]   E. Mikheev, B. D. Hoskins, D. B. Strukov, S. Stemmer, *Nat Commun* **2014**, *5*, 3990.

[15]   A. Baikalov, Y. Q. Wang, B. Shen, B. Lorenz, S. Tsui, Y. Y. Sun, Y. Y. Xue, Cw. Chu, *Appl Phys Lett* **2003**, *83*, 957.

[16]   S. Asanuma, H. Akoh, H. Yamada, A. Sawa, *Phys Rev B* **2009**, *80*, 235113.

[17]   A. Sawa, T. Fujii, M. Kawasaki, Y. Tokura, *Appl Phys Lett* **2004**, *85*, 4073.

[18]   Z. Fan, H. Fan, L. Yang, P. Li, Z. Lu, G. Tian, Z. Huang, Z. Li, J. Yao, Q. Luo, *J Mater Chem C Mater* **2017**, *5*, 7317.

[19]   Y. B. Nian, J. Strozier, N. J. Wu, X. Chen, A. Ignatiev, *Phys Rev Lett* **2007**, *98*, 146403.

[20]   D. Seong, M. Jo, D. Lee, H. Hwang, *Electrochemical and solid-state letters* **2007**, *10*, H168.

[21]   Y. S. Lin, F. Zeng, S. G. Tang, H. Y. Liu, C. Chen, S. Gao, Y. G. Wang, F. Pan, *J Appl Phys* **2013**, *113*.

[22]   R. Fors, S. I. Khartsev, A. M. Grishin, *Phys Rev B* **2005**, *71*, 045305.

[23]   T. Oka, N. Nagaosa, *Phys Rev Lett* **2005**, *95*, 266403.

[24]   M. J. Rozenberg, I. H. Inoue, M. J. Sanchez, *Phys Rev Lett* **2004**, *92*, 178302.

[25]   K. Baek, S. Park, J. Park, Y.-M. Kim, H. Hwang, S. H. Oh, *Nanoscale* **2017**, *9*, 582.

[26]   C.-H. Yang, J. Seidel, S. Y. Kim, P. B. Rossen, P. Yu, M. Gajek, Y.-H. Chu, L. W. Martin, M. B. Holcomb, Q. He, *Nat Mater* **2009**, *8*, 485.

[27]   A. Gruverman, D. Wu, H. Lu, Y. Wang, H. W. Jang, C. M. Folkman, M. Y. Zhuravlev, D. Felker, M. Rzchowski, C.-B. Eom, *Nano Lett* **2009**, *9*, 3539.

[28]	X. Zou, H. G. Ong, L. You, W. Chen, H. Ding, H. Funakubo, L. Chen, J. Wang, *AIP Adv* **2012**, *2*.

[29]	H. H. P. Garza, Y. Pivak, L. M. Luna, J. T. van Omme, R. G. Spruit, M. Sholkina, M. Pen, Q. Xu, in *2017 19th International Conference on Solid-State Sensors, Actuators and Microsystems (TRANSDUCERS)*, IEEE, **2017**, pp. 2155–2158.

[30]	V. Srot, R. Straubinger, F. Predel, P. A. van Aken, *Microscopy and Microanalysis* **2023**, *29*, 596.

[31]	P. Nukala, M. Ahmadi, Y. Wei, S. De Graaf, E. Stylianidis, T. Chakrabortty, S. Matzen, H. W. Zandbergen, A. Björling, D. Mannix, *Science* **2021**, *372*, 630.

[32]	H. Xiong, Z. Liu, X. Chen, H. Wang, W. Qian, C. Zhang, A. Zheng, F. Wei, *Science (1979)* **2022**, *376*, 491.

[33]	Y. Yang, S. Louisia, S. Yu, J. Jin, I. Roh, C. Chen, M. V Fonseca Guzman, J. Feijóo, P.-C. Chen, H. Wang, *Nature* **2023**, *614*, 262.

[34]	G. Zhu, B. A. Legg, M. Sassi, X. Liang, M. Zong, K. M. Rosso, J. J. De Yoreo, *Nat Commun* **2023**, *14*, 6300.

[35]	Y. Ling, T. Sun, L. Guo, X. Si, Y. Jiang, Q. Zhang, Z. Chen, O. Terasaki, Y. Ma, *Nat Commun* **2022**, *13*, 6625.

[36]	D. H. Pearson, C. C. Ahn, B. Fultz, *Phys Rev B* **1993**, *47*, 8471.

[37]	H. Tan, J. Verbeeck, A. Abakumov, G. Van Tendeloo, *Ultramicroscopy* **2012**, *116*, 24.

[38]	M. Varela, M. P. Oxley, W. Luo, J. Tao, M. Watanabe, A. R. Lupini, S. T. Pantelides, S. J. Pennycook, *Phys Rev B* **2009**, *79*, 085117.

[39]	E. Stoyanov, F. Langenhorst, G. Steinle-Neumann, *American Mineralogist* **2007**, *92*, 577.

[40]	Y. Shao, C. Maunders, D. Rossouw, T. Kolodiazhnyi, G. A. Botton, *Ultramicroscopy* **2010**, *110*, 1014.

[41]	A. Chen, W. Zhang, L. R. Dedon, D. Chen, F. Khatkhatay, J. L. MacManus-Driscoll, H. Wang, D. Yarotski, J. Chen, X. Gao, *Adv Funct Mater* **2020**, *30*, 2000664.

[42]    E. M. Bourim, S. Park, X. Liu, K. P. Biju, H. Hwang, A. Ignatiev, *Electrochemical and Solid-State Letters* **2011**, *14*, H225.

[43]    T. Choi, S. Lee, Y. J. Choi, V. Kiryukhin, S.-W. Cheong, *Science (1979)* **2009**, *324*, 63.

[44]    A. Chanthbouala, V. Garcia, R. O. Cherifi, K. Bouzehouane, S. Fusil, X. Moya, S. Xavier, H. Yamada, C. Deranlot, N. D. Mathur, *Nat Mater* **2012**, *11*, 860.

[45]    P. Roy, S. Kunwar, D. Zhang, D. Chen, Z. Corey, B.X. Rutherford, H. Wang, J.L. MacManus-Driscoll, Q. Jia, A. Chen, *Adv Elec Mater* 2022, *8*, 2101392.

[46]    S. K. Ravi, W. Sun, D. K. Nandakumar, Y. Zhang, S. C. Tan, *Sci Adv* **2018**, *4*, eaao6050.

[47]    B. Ofuonye, J. Lee, M. Yan, C. Sun, J.-M. Zuo, I. Adesida, *Semicond Sci Technol* **2014**, *29*, 095005.

[48]    D. Gu, S. K. Dey, P. Majhi, *Appl Phys Lett* **2006**, *89*.

[49]    Y. Zhuang, Y. Liu, H. Xia, Y. Li, X. Li, T. Li, *AIP Adv* **2022**, *12*.

[50]    L. P. B. Lima, H. F. W. Dekkers, J. G. Lisoni, J. A. Diniz, S. Van Elshocht, S. De Gendt, *J Appl Phys* **2014**, *115*.

[51]    P. Patsalas, S. Logothetidis, *J Appl Phys* **2001**, *90*, 4725.

[52]    V. Mansfeldova, M. Zlamalova, H. Tarabkova, P. Janda, M. Vorokhta, L. Piliai, L. Kavan, *The Journal of Physical Chemistry C* **2021**, *125*, 1902.

[53]    F. C. Marques, J. J. Jasieniak, *Appl Surf Sci* **2017**, *422*, 504.

[54]    A. M. Elsayed, M. Rabia, M. Shaban, A. H. Aly, A. M. Ahmed, *Sci Rep* **2021**, *11*, 17572.

[55]    C. Wang, K. Jin, Z. Xu, L. Wang, C. Ge, H. Lu, H. Guo, M. He, G. Yang, *Appl Phys Lett* **2011**, *98*.

[56]    D. Lee, S. H. Baek, T. H. Kim, J.-G. Yoon, C. M. Folkman, C. B. Eom, T. W. Noh, *Phys Rev B* **2011**, *84*, 125305.

[57]    G.-L. Yuan, J. Wang, *Appl Phys Lett* **2009**, *95*.

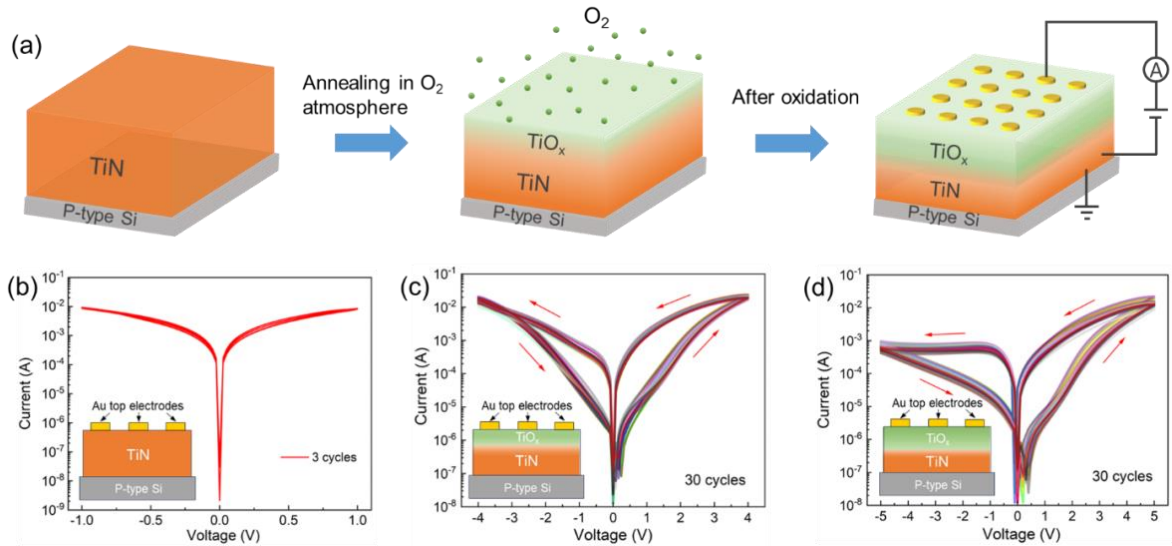

**Figure 1 Fabrication of Metal/TiO$_x$/TiN/Si memristor device and current-voltage (*I-V*) characteristics.** (a) Schematic of the TiN film oxidation and Au/TiO$_x$/TiN/Si device fabrication process. *I-V* switching curves of (b) Au/TiN/Si, (c) 1-minute and (d) 5-minute O$_2$ atmosphere annealed Au/TiO$_x$/TiN/Si devices.

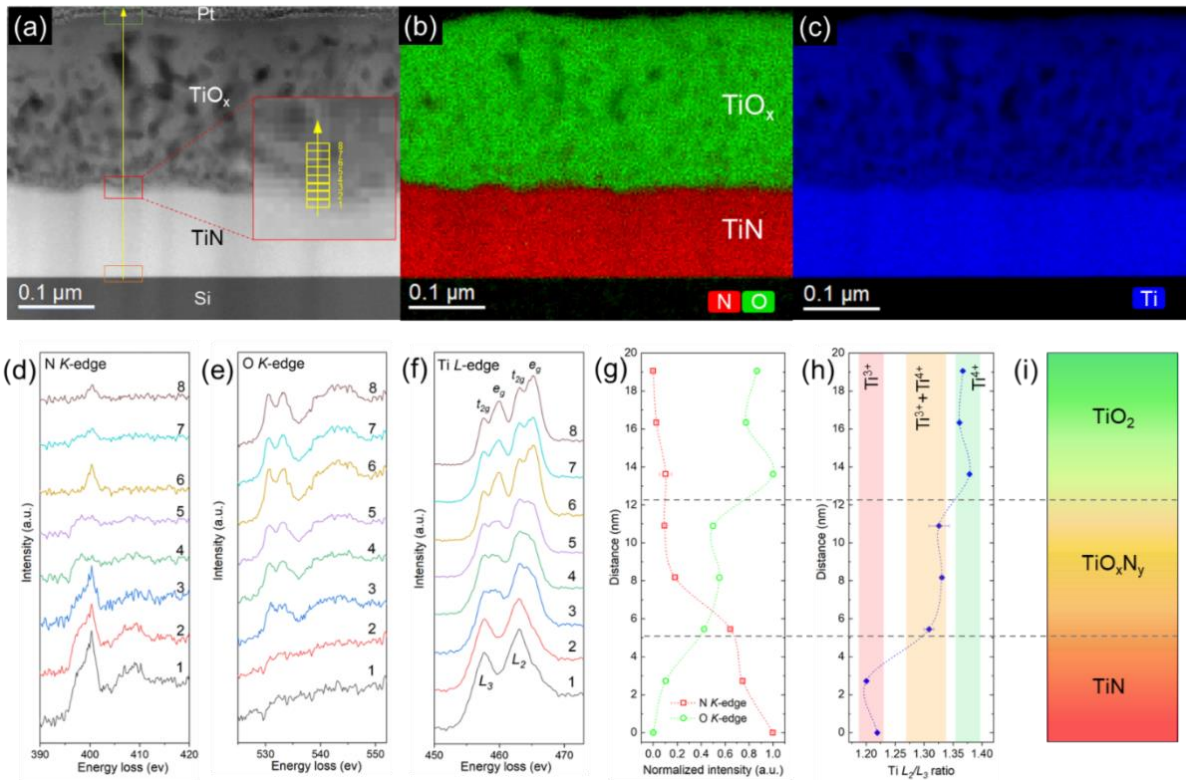

**Figure 2 HAADF-STEM imaging and EELS analysis of the Metal/TiO$_x$/TiN/Si memristor device.** (a) Cross-sectional HAADF-STEM image and STEM-EELS elemental maps of (b) N and O, and (c) Ti, respectively. (d-f) EELS intensities of the N-*K*, O-*K*, and Ti-*L$_{2,3}$* edges at the TiO$_x$/TiN interface marked as the enlarged inset in (a). (g) Normalized EELS intensities of N-*K* and O-*K* edges. (h) Calculated Ti-*L$_{2,3}$* intensity ratios, and (i) schematic of the TiO$_2$/TiO$_x$N$_y$/TiN structure based on (g) and (h).

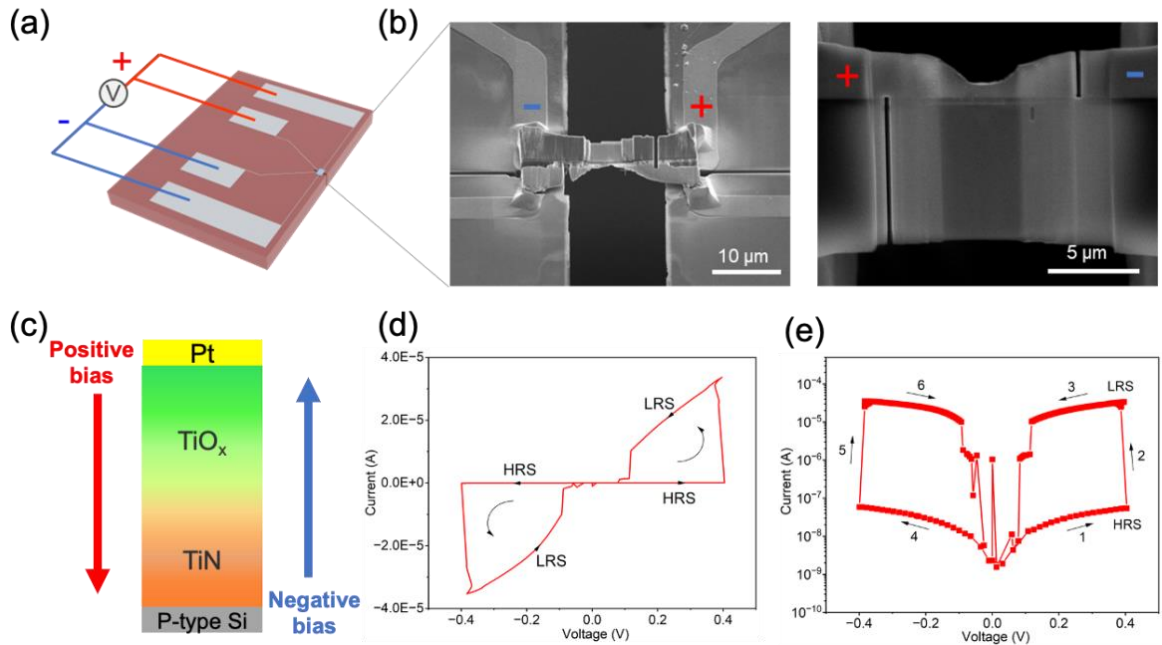

**Figure 3 *in situ* TEM biasing and *I-V* switching curves.** (a) Schematic of the MEMS-based E-chip and *in situ* biasing experiment. (b) SEM images showing the TEM lamella mounted on the E-chip after FIB milling and cleaning. (c) Schematic of the defined positive and negative bias directions applied to the vertical Pt/TiO$_x$/TiN/Si device. (d) *I-V* characteristic curves of the device obtained during the *in situ* TEM biasing experiment. (e) Log-scale *I-V* switching curves.

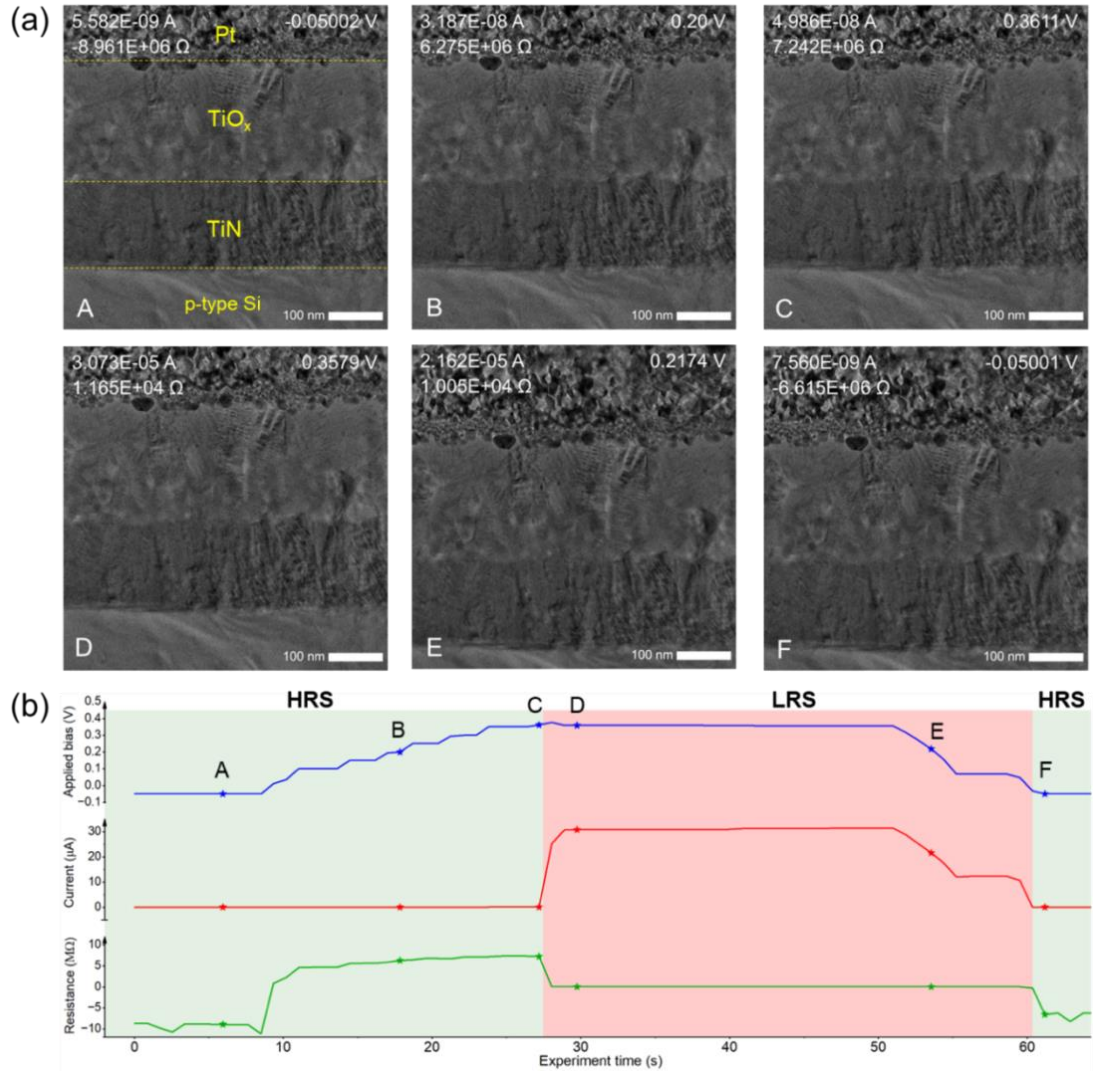

**Figure 4 *in situ* TEM images and electrical profiles.** (a) *in situ* TEM images (A to F) of the Pt/TiO$_x$/TiN/Si memristor acquired at different magnitude of electrical biases during the *in situ* biasing experiment. (b) Profiles of applied bias (V), current (μA), and resistance (MΩ) during the HRS → LRS → HRS switching process. Six data points labeled as A to F correspond to the electrical profiles of the TEM images in (a).

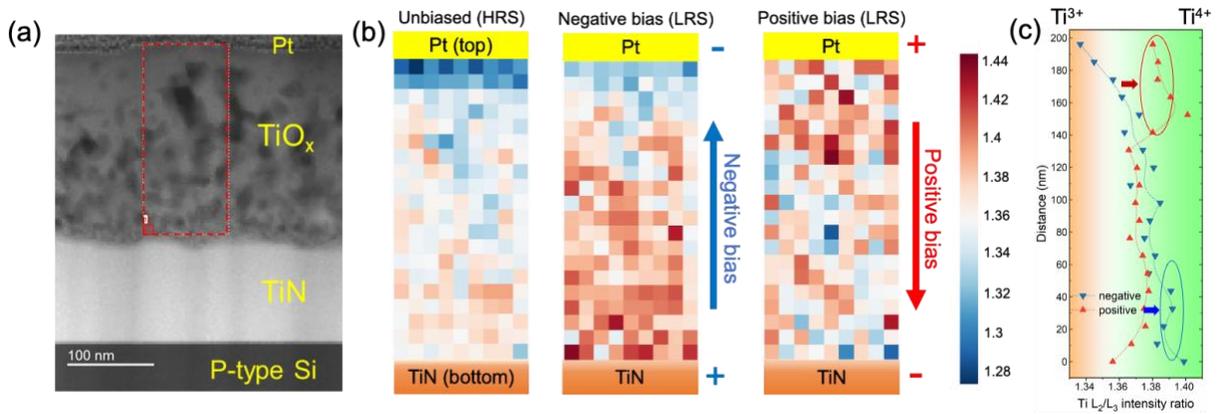

**Figure 5 ADF image and *in situ* EELS maps.** (a) Cross-sectional ADF image, (b) $L_2/L_3$ ratio maps derived from the *in situ* Ti-$L_{2,3}$ EELS spectra imaging (SI) at unbiased, negative and positive biased conditions. (c) Line profile of the averaged Ti $L_2/L_3$ intensity ratios for each row of the EELS map under negative and positive biases as shown in (b).

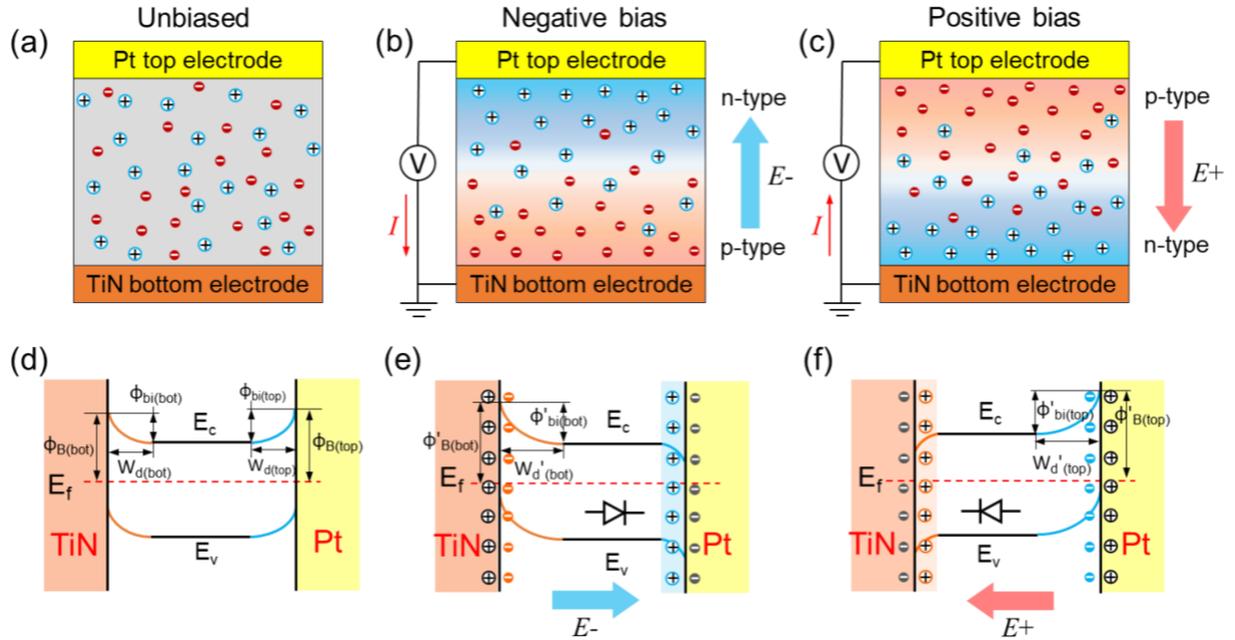

**Figure 6 The RS mechanism of the device.** (a-c) Schematic diagrams describing (a) the intrinsic electronic states, and motions of charge carriers under (b) negative and (c) positive electrical bias conditions, in which the inhomogeneous distribution of positive and negative charges results in the formation of electric-field-driven n-p or p-n junctions inside the device. (d-f) Schematic energy band diagrams of the Schottky-like barriers at the Pt/TiO$_2$ and TiO$_2$/TiN interfaces at the (d) unbiased, (e) negative and (f) positive biased conditions. In the diagrams, $\phi_{B(top)}$ and $\phi_{B(bot)}$ are the Schottky barriers, $W_{d(top)}$ and $W_{d(bot)}$ are the depletion width, and $\phi_{bi(top)}$ and $\phi_{bi(bot)}$ are the built-in potential barriers at the top Pt/TiO$_2$ and bottom TiO$_2$/TiN interfaces, respectively.

# Supporting Information

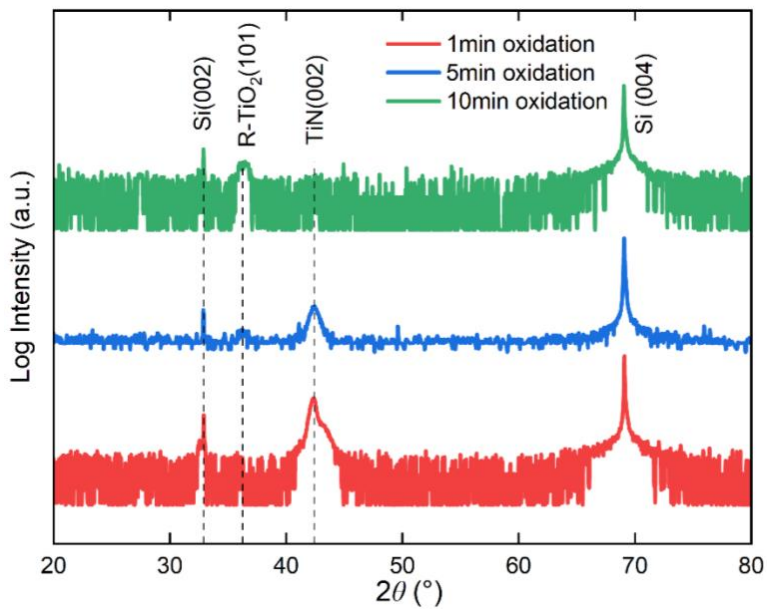

Figure S1 XRD $\theta$-$2\theta$ scans intensities of the TiOx/TiN/Si (001) films after 1-, 5-, and 10-min annealing in 200mTorr oxygen atmosphere.

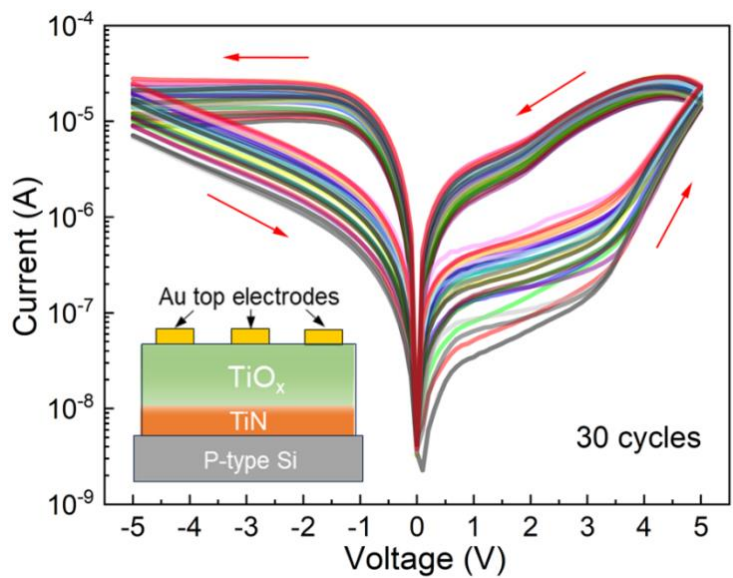

Figure S2 *I-V* switching curves of the 10-minute $O_2$ atmosphere annealed Au/TiO$_x$/TiN/Si device.

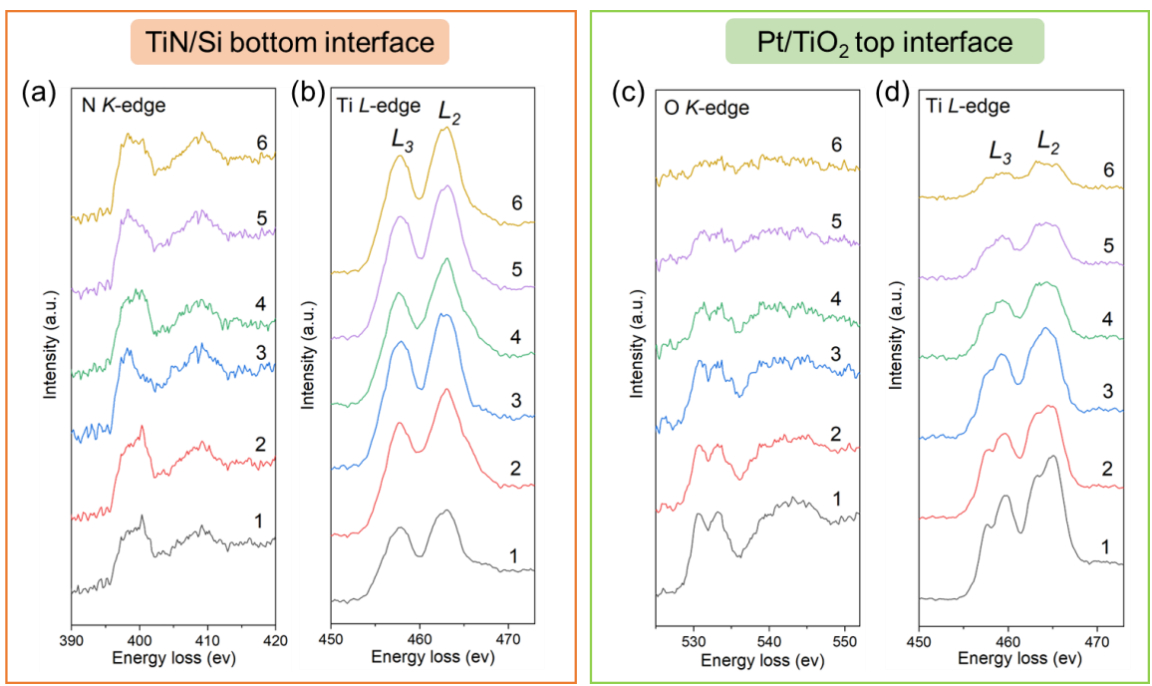

Figure S3 (a, b) EELS intensities of (a) *N-K*, and (b) *Ti-L$_{2,3}$* edges at the TiN/Si bottom interface. (c, d) EELS intensities of (c) *O-K*, and (d) *Ti-L$_{2,3}$* edges at the Pt/TiO$_2$ top interface.

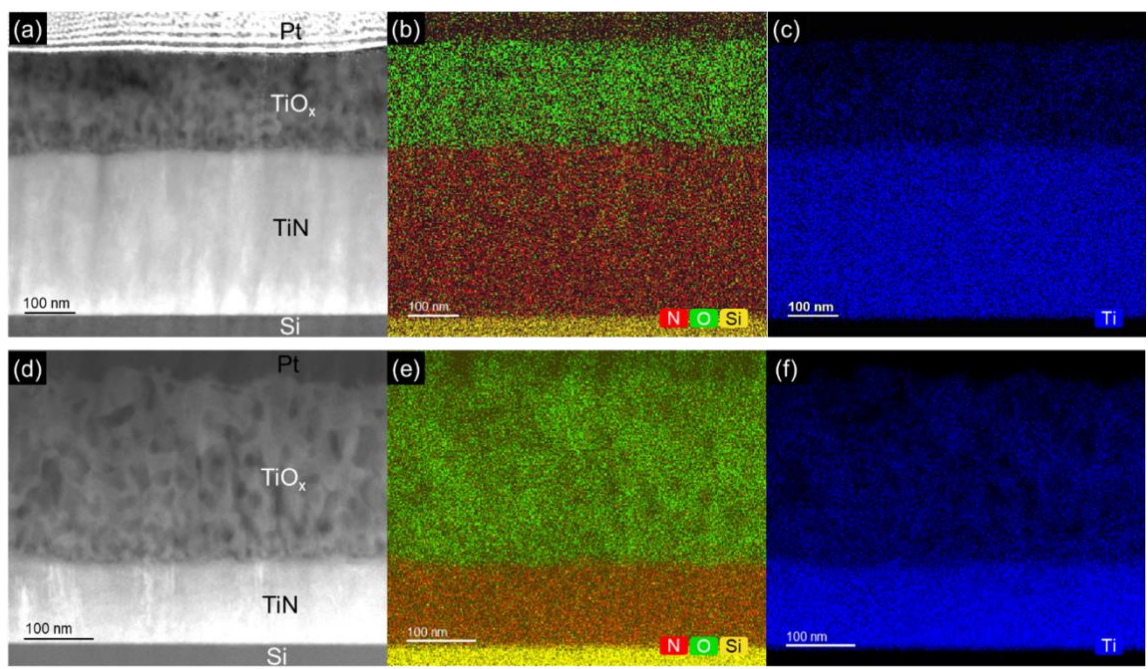

Figure S4 HAADF-STEM image and EDS element maps for N, O, Si, and Ti of the (a-c) 1-minute and (d-f) 5-minute O$_2$ atmosphere annealed TiO$_x$/TiN/Si samples.

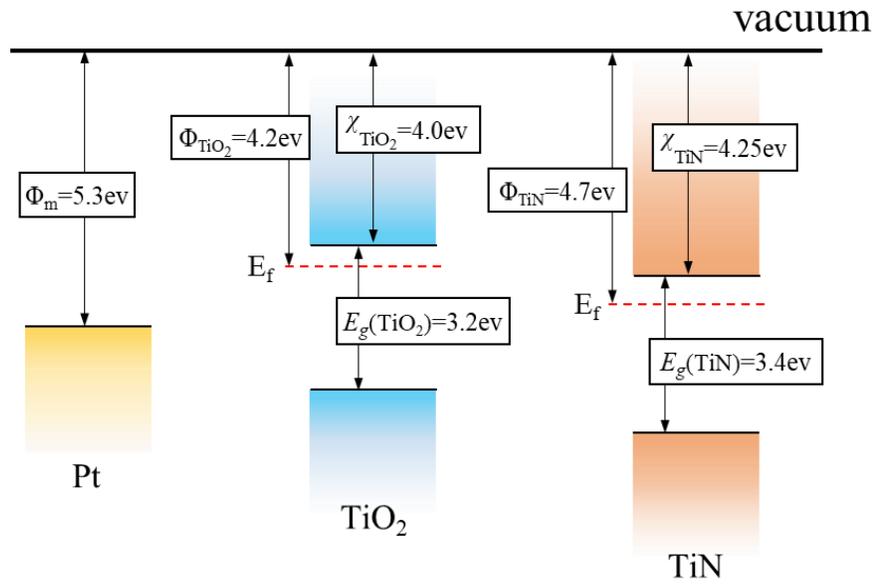

Figure S5 Energy band diagrams of Pt, TiO₂, and TiN. Where, $\phi_m$, $\phi_{TiO_2}$, and $\phi_{TiN}$ are the metal work functions of Pt, TiO₂, and TiN. $\chi_{TiO_2}$ and $\chi_{TiN}$ are the electron affinity values of TiO₂ and TiN. $E_g$(TiO₂) and $E_g$(TiN) are the band gaps of TiO₂ and TiN.

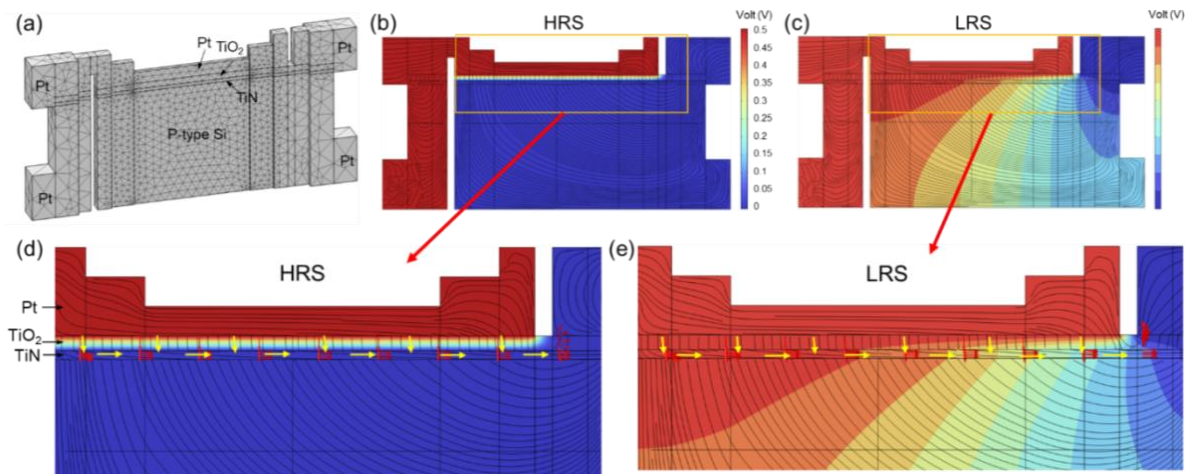

figure S6 (a) 3D geometry model built for COMSOL simulation. Simulated electric potential distribution maps with streamline curves for the device at (b) HRS and (c) LRS. Magnified images showing the electric potential and local electric field (current) directions at the RS interface of the device at (d) HRS and (e) LRS.

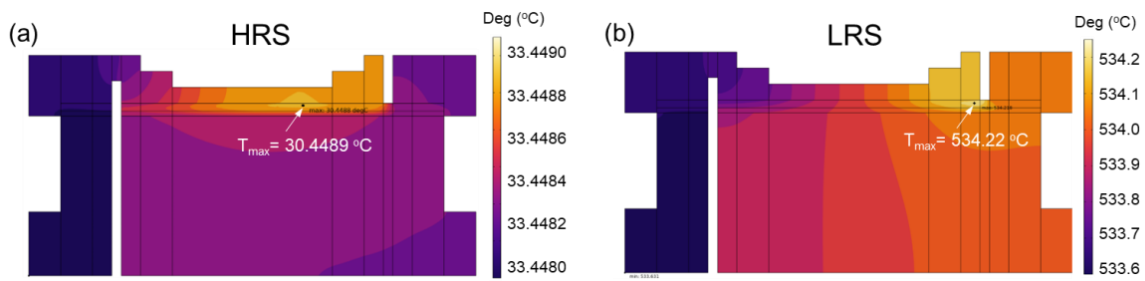

Figure S7 Electro-thermal simulation results showing the temperature distribution map on the device at (a) HRS with current of 44.9 nA and (b) LRS with current of 20.7 µA. Here, the heat transfer coefficient ($h$) is set to 10 W/m$^2$K, and surface emissivity ($\varepsilon$) is set to 0.8.

Table S1    Materials physical parameters used in the electro-thermal simulation.

| Heat transfer coefficient ($h$) (W/m²K) | Surface emissivity ($\varepsilon$) | $T_{max}$ (°C) |
|---|---|---|
| 5 | 0.5 | 654.03 |
| 5 | 0.8 | 556.82 |
| 10 | 0.8 | 534.22 |
| 15 | 0.8 | 511.81 |
| 20 | 0.8 | 489.84 |
| 20 | 1 | 456.51 |

For the Joule heating effect simulation, the terminal voltage was set to 0.5 V, and electric currents flowing through the device was set to 44.9 nA at HRS, and 20.7 µA at LRS. Different heat transfer coefficient and surface emissivity values were used to simulate the $T_{max}$ inside the device at LRS during RS.

It can be seen that the simulated $T_{max}$ values reduce with the increase of heat transfer coefficient ($h$) and surface emissivity ($\varepsilon$). As for most of the materials, $h$ is about 5-15, and $\varepsilon$ is less than 1.0 (1.0 is for blackbody), the estimated $T_{max}$ can reach to 500 °C or even higher.